\documentclass{article}
\usepackage{amsmath}
\usepackage{amssymb}

\input{tcilatex}

\begin{document}

\title{Couplings of the $\eta$ and $\eta^{\prime}$ Mesons to the Nucleon}
\author{N. F. Nasrallah \\
Faculty of Science, Lebanese University\\
Tripoli, Lebanon\\
nsrallh@ul.edu.lb\\
\linebreak }
\date{}
\maketitle

\begin{abstract}
The couplings of the $\eta$ and $\eta^{\prime}$ mesons to the nucleon are
obtained from the $U_{A}(1)$ Goldberger-Treiman relation. The chiral
symmetry breaking corrections are very large and bring the calculated values
of the coupling constants $G_{\eta NN}$ and $G_{\eta^{\prime}NN}$ close to
values obtained from potential models.
\end{abstract}

The coupling constants of the nucleon to the $\pi $ and to the $\eta $ and $%
\eta ^{\prime }$ mesons enter in the study of low energy hadronic physics
especially in the description of nucleon-nucleon scattering data \cite{Grein}%
, $\eta $ photoproduction \cite{Ajaka}, in the estimates of the electric
dipole moment of the neutron \cite{Ohta} or of the proton-neutron mass
difference \cite{Schechter}. While the value of $G_{\pi NN}$ is known with
reasonable precision, the values of $G_{\eta NN}$ and $G_{\eta ^{\prime }NN}$
are not \cite{Dumbrajs}. For the pion, one has the well known
Goldberger-Treiman relation \cite{Goldberger}. 
\begin{equation}
f_{\pi }.G_{\pi NN}=\sqrt{2}m_{N}G^{(3)}  \label{1}
\end{equation}%
Which involves the pion decay constant $f_{\pi }=.0924$\textrm{Gev} and the
renormalized axial-vector coupling constant $G^{(3)}=(1.267\pm .004)/\sqrt{2}
$ defined through 
\begin{equation}
\langle N(p,s)\left\vert A_{\mu }^{(3)}\right\vert N(p,s)\rangle
=G^{(3)}s_{\mu }/\sqrt{2}
\end{equation}%
at vanishing momentum transfer.

The Goldberger-Treiman relation is satisfied to a very good accuracy \cite%
{Nasrallah}. This is not surprising because the approximations used to
obtain eq. (\ref{1}) involve an extrapolation in momentum transfer squared
from $0$ to $m_{\pi}^{2}$ which is a very small quantity on the hadronic
scale.

The straightforward generalisation of eq. (\ref{1}) to the $\eta$ and $%
\eta^{\prime}$ channels was considered in \cite{Feldman} who obtained 
\begin{equation}
G_{\eta NN}=3.4\pm .5\ \ \ \ \ \ \ \ \ \ \ ,\ \ \ \ \ \ \ \ \ \ \
G_{\eta^{\prime}NN}=1.4\pm 1.1
\end{equation}
values which differ considerably from ones obtained in potential models \cite%
{Dumbrajs}: 
\begin{equation}
G_{\eta NN}=6.8\ \ \ \ \ \ \ \ \ \ \ ,\ \ \ \ \ \ \ \ \ \ \
G_{\eta^{\prime}NN}=7.3  \label{4}
\end{equation}
The $\eta$ and $\eta ^{\prime}$ mesons are however very heavy and the
extrapolation from the chiral limit to the values of the physical masses
potentially introduces large corrections.

It is the purpose of this work to undertake such an extrapolation. This will
be done under the sole assumption that corrections to chiral symmetry
breaking arise mainly from the contributions of intermediate states in the
isoscalar $0^{-}$ continuum with invariant mass squared in the interval
extending from $1.5\mathrm{Gev}^{2}$ to $3.5\mathrm{Gev}^{2}$ which includes
the resonances $\eta(1295)$, $\eta(1405),$ $\eta(1475)$ as well as the newly
discovered $X(1835)$ which couples strongly to the nucleon \cite{Abikim}.

It will be seen that, when chiral symmetry breaking is taken into account,
the $U_{A}(1)$ Goldberger-Treiman relation yields for $G_{\eta NN}$ and$%
G_{\eta^{\prime}NN}$ values close to the ones appearing in eq. (\ref{4}).

The isoscalar components of the octet of axial-vector currents couple to
the\ $\eta $ and $\eta ^{\prime }$ mesons: 
\begin{eqnarray}
\langle 0\left\vert A_{\mu }^{(8)}\right\vert \eta (p)\rangle &=&if_{8}\cos
\theta \text{ }p_{\mu }\ \ \ \ \ \ \ \ ,\ \ \ \ \ \ \ \ \langle 0\left\vert
A_{\mu }^{(8)}\right\vert \eta ^{\prime }(p)\rangle =if_{8}\sin \theta \text{
}p_{\mu }  \notag \\
\ \langle 0\left\vert A_{\mu }^{(0)}\right\vert \eta (p)\rangle
&=&-if_{0}\sin \theta \text{ }p_{\mu }\ \ \ \ \ \ \ \ ,\ \ \ \ \ \ \ \
\langle 0\left\vert A_{\mu }^{(0)}\right\vert \eta ^{\prime }(p)\rangle
=if_{0}\cos \theta \text{ }p_{\mu }
\end{eqnarray}%
$\theta $ is the singlet-octet mixing angle. In the $SU(3)$ limit $%
f_{8}=f_{\pi }$. The axial-vector currents are expressed in terms of quark
fields 
\begin{eqnarray}
A_{\mu }^{(8)} &=&\frac{1}{2\sqrt{3}}(\overline{u}\gamma _{\mu }\gamma _{5}u+%
\overline{d}\gamma _{\mu }\gamma _{5}d-2\overline{s}\gamma _{\mu }\gamma
_{5}s)  \notag \\
A_{\mu }^{(0)} &=&\frac{1}{\sqrt{6}}(\overline{u}\gamma _{\mu }\gamma _{5}u+%
\overline{d}\gamma _{\mu }\gamma _{5}d+\overline{s}\gamma _{\mu }\gamma
_{5}s)
\end{eqnarray}

A two angle description for the octet and singlet components, $\theta_{8}$
and $\theta_{0}$, has more recently been advocated \cite{Leutwyler}. We
shall see that this description does not alter our numerical analysis.

When the divergence of the currents is taken, the singlet component picks up
a gluon anomaly term 
\begin{eqnarray}
D^{(8)} &=&\partial _{\mu }A_{\mu }^{(8)}=\frac{i}{\sqrt{3}}(m_{u}\overline{u%
}\gamma _{5}u+m_{d}\overline{d}\gamma _{5}d-2m_{s}\overline{s}\gamma _{5}s) 
\notag \\
D^{(0)} &=&\partial _{\mu }A_{\mu }^{(0)}=i\sqrt{\frac{2}{3}}(m_{u}\overline{%
u}\gamma _{5}u+m_{d}\overline{d}\gamma _{5}d+m_{s}\overline{s}\gamma _{5}s)+Q
\end{eqnarray}%
Where 
\begin{equation}
Q=\frac{1}{\sqrt{6}}\text{ }\frac{3\alpha _{s}}{4\pi }G_{\mu \nu }\widetilde{%
G}^{\mu \nu }
\end{equation}%
$G_{\mu \nu }$ being the gluonic field strength tensor and $\widetilde{G}%
_{\mu \nu }=\frac{1}{2}\epsilon _{\mu \nu \rho \sigma }G^{\rho \sigma }$
it's dual.

Consider the matrix element 
\begin{equation}
\langle N(p^{\prime })\left\vert D^{(8)}/f_{8}\right\vert N(p)\rangle =\Pi
_{8}(t)\text{ }\overline{U}\gamma _{5}U
\end{equation}%
with $t=(p-p^{\prime })^{2}$. We have 
\begin{equation}
\Pi _{8}(t=0)=\sqrt{2}m_{N}G^{(8)}/f_{8}
\end{equation}%
where $G^{(8)}$ is the isoscalar octet axial -vector coupling constant
related to $G^{(3)}$ through the $F/D$ ratio obtained from Hyperon decay, $%
G^{(8)}=.24$.

In the low energy domain, $\Pi _{8}(t)$ is dominated by the $\eta $ and $%
\eta ^{\prime }$ poles, i.e. 
\begin{equation}
\Pi _{8}(t)=\frac{m_{\eta }^{2}\cos \theta }{m_{\eta }^{2}-t}\text{ }G_{\eta
NN}+\frac{m_{\eta ^{\prime }}^{2}\sin \theta }{m_{\eta ^{\prime }}^{2}-t}%
\text{ }G_{\eta ^{\prime }NN}+other~terms
\end{equation}%
$\Pi _{8}(t)$ is an analytic function in the complex $t$-plane with poles at
\ $t=m_{\eta }^{2}$ and $t=m_{\eta ^{\prime }}^{2}$ and a cut along the
positive $t$-axis which starts at the $(\eta +2\pi )$ threshold.

An immediate consequence of Cauchy's theorem is%
\begin{equation}
\frac{1}{2\pi i}\dint\limits_{C}\frac{dt}{t}\text{ }\Pi _{8}(t)=\frac{1}{%
2\pi i}\dint\limits_{th}^{R}\frac{dt}{t}\text{ }Disc\Pi _{8}(t)+\frac{1}{%
2\pi i}\doint \frac{dt}{t}\text{ }\Pi _{8}(t)=\sqrt{2}m_{N}G^{(8)}/f_{8}-G_{%
\eta NN}\cos \theta -G_{\eta ^{\prime }NN}\sin \theta  \label{12}
\end{equation}%
where $C$ \ is a closed contour consisting of two straight lines immediately
above and below the cut which starts effectively at $t_{1}=(m_{\eta ^{\prime
}}+2m_{\pi })^{2}$ and a circle of large radius in the complex $t-$plane $%
(R\simeq 4-5GeV^{2})$. The contribution of the non-resonant threshold part
of the continuum extending from $t_{0}=(m_{\eta }+2m_{\pi })^{2}$ to $t_{1}$
is small compared to the one of the $\eta $ and $\eta ^{\prime }$ mesons and
is neglected. Treating the $SU(3)$ singlet amplitude $\Pi _{0}(t)$ in a
similar fashion gives 
\begin{equation}
\frac{1}{2\pi i}\dint\limits_{C}\frac{dt}{t}\text{ }\Pi _{0}(t)=\sqrt{2}%
m_{N}G^{(0)}/f_{0}+G_{\eta NN}\sin \theta -G_{\eta ^{\prime }NN}\cos \theta
\label{13}
\end{equation}%
If the left hand sides of eqs. (\ref{12}) and (\ref{13}) , which represent
the corrections due to the contributions of the $0^{-}$ continuum were
negligible,these equations would yield for the coupling constants $G_{\eta
NN}$ and $G_{\eta ^{\prime }NN}$ the values obtained in ref. \cite{Feldman}.
The neglect of the contributions of the continuum is however not justified,
they are of order $m_{\eta }^{2}$ and $m_{\eta ^{\prime }}^{2}$ which are
not small quantities on the hadronic scale.

In order to estimate these contributions, use will be made of the modified
integral $\frac{1}{2\pi i}\dint_{C}dt(\frac{1}{t}-a_{0}-a_{1}$ $t)\Pi
_{8}(t) $ where $a_{0}$ and $a_{1}$ are so far arbitrary constants. The
residue theorem yields%
\begin{eqnarray}
&&G_{\eta NN}\cos \theta (1-a_{0}m_{\eta }^{2}-a_{1}m_{\eta }^{4})+G_{\eta
^{\prime }NN}\sin \theta (1-a_{0}m_{\eta ^{\prime }}^{2}-a_{1}m_{\eta
^{\prime }}^{4})-\sqrt{2}m_{N}G^{(8)}/f_{8}  \notag \\
&=&\frac{1}{2\pi i}\dint\limits_{th}^{R}dt(\frac{1}{t}-a_{0}-a_{1}\text{ }%
t)Disc\Pi _{8}(t)+\frac{1}{2\pi i}\doint dt(\frac{1}{t}-a_{0}-a_{1}\text{ }%
t)\Pi _{8}^{QCD}(t)  \label{14}
\end{eqnarray}

In the first of the integrals above, over the cut, the main contribution is
expected to arise from the interval $I:1.5\mathrm{Gev}^{2}\leq t\leq 3.5%
\mathrm{Gev}^{2}$ where the PDG \cite{Eidelman} lists three $0^{-}$
resonances $\eta (1295)$, $\eta (1405)$ and $\eta (1475)$ in addition to the
newly discovered $X(1835)$ \cite{Abikim}\ . The constants $a_{0}$ and $a_{1}$
are now chosen so as to practically annihilate the kernel $(\frac{1}{t}%
-a_{0}-a_{1}.t)$ on the interval $I$. The choice.

\begin{equation}
a_{0}=.879\mathrm{Gev}^{-2}\ \ \ \ \ \ \ \ \ \ \ ,\ \ \ \ \ \ \ \ \ \
a_{1}=-.177\mathrm{Gev}^{-4}
\end{equation}%
obtained by a least square fit, reduces the integrand to a few percent of
it's initial value on $I$. This allows the neglect of the integral over the
cut.

In the second integral,over the circle, $\Pi _{8}(t)$ has been replaced by $%
\Pi _{8}^{QCD}(t)$ an approximation which is good except possibly for a
small region in the vicinity of \ the positive real $t$ axis.This integral
is likewise negligible even though the kernel $(\frac{1}{t}-a_{0}-a_{1}.t)$
is no longer small because $\Pi _{8}^{QCD}(t)$ itself is small , $\Pi
_{8}^{QCD}(t)\thicksim m_{u,d}\langle \overset{-}{q}q\rangle .$

This can be seen by using the method of $QCD$ sum-rules \cite{RRY}. Because
the nucleon currents do not involve the strange quark $s$ , $\Pi
_{8}^{QCD}(t)$ can be obtained from a study of the three point function

\begin{equation*}
A(p,p^{\prime},q)=\frac{i}{\sqrt{3}f_{8}}\int \int
dxdye^{ipx}e^{ip^{\prime}y}\left\langle 0\right\vert T\eta(x)J_{8}(0)%
\overline{\eta}(y)\left\vert 0\right\rangle
\end{equation*}

where $\eta$ is the nucleon current with coupling $\lambda_{N}$ to the
nucleon, $J_{8}=m_{u}\overline{u}\gamma_{5}u+m_{d}\overline{d}\gamma_{5}d$.

$A(p,p^{\prime},q)$ has a double nucleon pole $(p=p^{\prime})$

\begin{equation}
A=i\gamma.q\gamma _{5}\left[ \frac{\lambda _{N}^{2}m_{N}^{2}}{%
(p^{2}-m_{N}^{2})^{2}}\Pi_{8}(q^{2})+...\right] +other~tensor~structures
\label{16}
\end{equation}

In the deep euclidean region, $A^{QCD}$ is given by the operator product
expansion.

\begin{equation}
A^{QCD}=c_{u}(p,p^{\prime},q)\left\langle 0\left\vert m_{u}\overline{u}%
u\right\vert 0\right\rangle +c_{d}(p,p^{\prime },q)\left\langle 0\left\vert
m_{d}\overline{d}d\right\vert 0\right\rangle  \label{17}
\end{equation}

$\Pi _{8}^{QCD}$ is obtained by extrapolating to the nucleon mass shell,
i.e. by taking the Borel (Laplace) transform of eqs. (\ref{16}) and (\ref{17}%
) with respect to the variable $p^{2}$ while keeping $t=q^{2}$ large and
negative. Identifying terms shows that $\Pi_{8}^{QCD}(t)$ is proportional to 
$\left\langle 0\left\vert m_{q}\overline{q}q\right\vert 0\right\rangle.$

Note that the contribution of the gluonic term $Q$ (which contributes only
to $\Pi_{0}(t)$) on the circle is not necessarily small. For this reason $%
\Pi_{0}(t)$ will not be used in this calculation.

With our choice of the coefficients $a_{0}$ and $a_{1}$ the integrals on the
r.h.s. of eq. (\ref{14}) thus becomes negligible and we have 
\begin{equation}
G_{\eta NN}\text{ }\cos \theta \text{ }(1-a_{0}m_{\eta }^{2}-a_{1}m_{\eta
}^{4})+G_{\eta ^{\prime }NN}\text{ }\sin \theta \text{ }(1-a_{0}m_{\eta
^{\prime }}^{2}-a_{1}m_{\eta ^{\prime }}^{4})=\sqrt{2}m_{N}G^{(8)}/f_{8}
\label{18}
\end{equation}

The chiral symmetry breaking corrections now show up in the coefficients of
the coupling constants.

More information is obtained by reiterating the same procedure with a
quadratic fit to $\frac{1}{t}$ i.e. use the integral $\frac{1}{2\pi i}%
\int_{c}dt(\frac{1}{t}-b_{0}-b_{1}t-b_{2}t^{2})\Pi _{8}(t)$ with fit
coefficients 
\begin{equation}
b_{0}=1.380\mathrm{Gev}^{-2},\ \ \ \ \ \ \ \ \ \ \ b_{1}=-.607\mathrm{Gev}%
^{-4},\ \ \ \ \ \ \ \ \ \ \ b_{2}=.085\mathrm{Gev}^{-6}
\end{equation}

This yields 
\begin{equation}
G_{\eta NN}\text{ }\cos \theta \text{ }(1-b_{0}m_{\eta }^{2}-b_{1}m_{\eta
}^{4}-b_{2}m_{\eta }^{6})+G_{\eta ^{\prime }NN}\text{ }\sin \theta \text{ }%
(1-b_{0}m_{\eta ^{\prime }}^{2}-b_{1}m_{\eta ^{\prime }}^{4}-b_{2}m_{\eta
^{\prime }}^{6})=\sqrt{2}m_{N}G^{(8)}/f_{8}  \label{20}
\end{equation}

Numerically eqs. (\ref{18}) and (\ref{20})\ boil down to 
\begin{eqnarray}
.752\text{ }G_{\eta NN}\cos \theta +.341\text{ }G_{\eta ^{\prime }NN}\text{ }%
\sin \theta  &=&\sqrt{2}m_{N}\text{ }G^{(8)}/f_{8}  \notag \\
.643\text{ }G_{\eta NN}\cos \theta +.178\text{ }G_{\eta ^{\prime }NN}\text{ }%
\sin \theta  &=&\sqrt{2}m_{N}\text{ }G^{(8)}/f_{8}  \label{21}
\end{eqnarray}

The coupling constant $f_{8}$ is not measured experimentally. It can be
obtained from the two photon decay rate of the $\eta $ and $\eta ^{\prime }$
mesons 
\begin{eqnarray}
\Gamma (\eta &\rightarrow &2\gamma )=\frac{\alpha ^{2}m_{\eta }^{3}}{192\pi
^{3}}\text{ }(\frac{\cos \theta }{f_{8}}-\frac{2\sqrt{2}\sin \theta }{f_{0}}%
)^{2}\text{ }(1+\Delta _{\eta })^{2}  \notag \\
\Gamma (\eta ^{\prime } &\rightarrow &2\gamma )=\frac{\alpha ^{2}m_{\eta
^{\prime }}^{3}}{192\pi ^{3}}\text{ }(\frac{\sin \theta }{f_{8}}+\frac{2%
\sqrt{2}\cos \theta }{f_{0}})^{2}\text{ }(1+\Delta _{\eta ^{\prime }})^{2}
\label{22}
\end{eqnarray}%
$\Delta _{\eta }$ and $\Delta _{\eta ^{\prime }}$ represent chiral symmetry
breaking correction factors which are often neglected in the litterature but
which have been evaluated in ref. \cite{Nasrallah2} and found to be large:%
\begin{equation}
\Delta _{\eta }=.77\ \ \ \ \ \ \ \ \ \ \ and\ \ \ \ \ \ \ \ \ \ \ \Delta
_{\eta ^{\prime }}=6.0  \label{23}
\end{equation}%
The corresponding values of $f_{8}$ ,$f_{8}=1.27f_{\pi },1.55$ $f_{\pi }$\
for \ $\theta =-18.5^{0},-30.5^{0}$ respectively, are obtained from the
measured values of the decay rates.

The coupling constants $G_{\eta NN}$ and $G_{\eta ^{\prime }NN}$ thus come
out as a function of the mixing angle $\theta .$

Traditionally, analyses based on the Gell-Mann-Okubo mass formula led to the
adoption of a small value for the mixing angle, $\theta \backsimeq -10^{0}$.
Subsequent study of the axial-anomaly generated decays $\eta ,\eta ^{\prime
}\rightarrow 2\gamma $ and measurement of the decay rates led to the
adoption of larger values for $\theta $, $-25^{0}\lesssim \theta \lesssim
20^{0}\ $\cite{Donohue}. Recently \cite{Nasrallah3}\ the corrections due to
chiral symmety breaking \ to the Gell-Mann-oakes-Renner relation,to the
calculation of the decay rate $\Gamma (\eta \rightarrow 3\pi )$ and to
radiative decays involving $\eta $ and $\eta ^{\prime }$were evaluated.It
was found that good agreement with experiment and numerical stability is
obtained for values of $\theta $ in the range $-30.5^{0}\leq \theta \leq
-18.5^{0}.$

Varying $\theta $ within these limits yields%
\begin{eqnarray}
G_{\eta NN} &=&5.45\ \ \ \ \ \ \ ,\ \ \ \ \ \ \ G_{\eta ^{\prime }NN}=10.90\
\ \ \ \ \ \ for\ \theta =-18.5^{0}  \notag \\
G_{\eta NN} &=&4.95\ \ \ \ \ \ \ ,\ \ \ \ \ \ \ G_{\eta ^{\prime }NN}=5.60\
\ \ \ \ \ \ for\ \theta =-30.5^{0}
\end{eqnarray}

In order to get an estimate of the error and to check the stability of the
calculation a cubic fit to $\frac{1}{t}$ can be used, i.e. take for kernel
in the dispersion integral $(1/t-c_{0}-c_{1}t-c_{2}t^{2}-c_{3}t^{3})$ with%
\begin{eqnarray}
c_{0} &=&1.85GeV^{-2}\ \ \ \ \ ,\ \ \ \ \ \ \ c_{1}=-1.263GeV^{-4}  \notag \\
c_{2} &=&.375GeV^{-6}\ \ \ \ \ \ ,\ \ \ \ \ \ \ c_{3}=-.041GeV^{-8}
\end{eqnarray}%
resulting in

\begin{equation}
.548\text{ }G_{\eta NN}\cos \theta +.103\text{ }G_{\eta ^{\prime }NN}\sin
\theta =\sqrt{2}m_{N}G^{(8)}/f_{8}
\end{equation}

For the values of the coupling constants $G_{\eta NN}$ and $G_{\eta ^{\prime
}NN}$ we have obtained and for the range of variation of $\theta $
considered, the discrepancy between the two sides of the equation above
amounts to $2\%-8\%$ . This shows that the calculation is stable and gives
an idea of the magnitude of the errors involved.

In the two angle description \cite{Leutwyler} the mixing angles $\theta _{8}$
and $\theta _{0}$ accompany the octet and singlet components. Because only
the octet component has been used in this calculation the corresponding
results hold with $\theta $ replaced by $\theta _{8}$ . \ The values $\theta
_{8}=-21.2^{0}$, $f_{8}=1.26f_{\pi }$ resulting from the analysis of ref. 
\cite{Feldman} which correspond to $\theta =-21.2^{0}$, $f_{8}=1.31f_{\pi }$
in my case yield practically identical results for the coupling constants.\
\ \ \ \ \ \ \ \ \ \ \ \ \ \ \ \ \ \ \ \ \ \ \ \ \ \ \ \ \ \ \ \ \ 

It is seen that the values obtained for the coupling constants are close to
those appearing in eq. (\ref{4}) which are obtained in potential models.

Chiral symmetry breaking thus bridges the gap between values of the coupling
constants obtained from the Goldberger-Treiman relation \ and those obtained
from potential models.

It is finally to be noted that while $G_{\eta NN}$, as well as all physical
quantities studied in \cite{Nasrallah3} are quite insensitive to the exact
value of the mixing angle, this is not the case for $G_{\eta ^{\prime }NN}.$%
\pagebreak


\begin{thebibliography}{99}
\bibitem{Grein} W. Grein and P. Kroll, Nucl. Phys. \textbf{A338} (1980),332;
M.M. Nagels, T.A. Rijken and J.J. de Swart, Phys. Rev. \textbf{D17} (1978),
768.

V.G.J. Stoks and T.A. Rijken, Nucl. Phys. \textbf{A613} (1997), 311.

\bibitem{Ajaka} J. Ajaka et al., GRAAL collaboration, Phys. Rev. Lett. 
\textbf{81} (1998), 1797.

\bibitem{Ohta} K. Kawarabayashi and N. Ohta, Nucl. Phys. \textbf{B175}
(1980), 1789.

\bibitem{Schechter} J. Schechter, A. Subbaraman and H. Weigel, Phys. Rev. 
\textbf{D48} (1993), 339.

\bibitem{Dumbrajs} O. Dumbrajs et al., Nucl. Phys. \textbf{B216} (1983), 277
and references therein.

\bibitem{Goldberger} M. Goldberger and S.B. Treiman, Phys. Rev. \textbf{110}
(1958), 1178.

\bibitem{Nasrallah} D.V. Bugg and M.D. Scadron, hep-ph/0312346; N.F.
Nasrallah, Phys.Rev. \textbf{D62} (2000), 036006 and references therein.

\bibitem{Feldman} T. Feldman, Int. J. Mod. Phys. \textbf{A15} (2000), 159;\
hep-ph/9907491\ and references therein.

\bibitem{Abikim} M. Abikim et al., BES collaboration, hep-ex/0508025; J.Z.
Bai et al., BES collaboration, Phys. Rev. Lett. \textbf{91} (2003) 022001.

\bibitem{Leutwyler} H. Leutwyler and R. Kaiser, Eur. Phys. J. C17 (2000),
623; P. Kroll and T. Feldmann, Phys. Scripta T99 (2002) 13.

\bibitem{Eidelman} S. Eidelman et al., PDG, Phys. Lett. \textbf{B592}
(2004), 1.

\bibitem{RRY} For a general review see L.J. Reinders, H. Rubinstein and
S.Yazaki, Phys. Rep. 127 (1985), 1

\bibitem{Nasrallah2} N.F. Nasrallah, Phys. Rev. \textbf{D66} (2002) 076012.

\bibitem{Donohue} J.F. Donoghue, B.R. Holstein and Y.C.R. Lin, Phys. Rev.
Lett. \textbf{55} (1985) 2766 ; F.J. Gilman and R. Kauffman, Phys. Rev. 
\textbf{D36} (1987)2761 and references therein.

\bibitem{Nasrallah3} N.F. Nasrallah, Phys. Rev. \textbf{D70} (2004) 116001.
\end{thebibliography}
\end{document}